# Energy Efficient Power Allocation in Massive MIMO Systems with Mismatch Channel Estimation Error

Mahdi Nangir
Electrical and Computer Engineering Department
University of Tabriz
Tabriz, Iran
nangir@tabrizu.ac.ir

Abdolrasoul Sakhaei Gharagezlou
Electrical and Computer Engineering Department
University of Tabriz
Tabriz, Iran
abdolrasoulsakhaei@gmail.com

Nima Imani
Electrical and Computer Engineering Department
University of Tabriz
Tabriz, Iran
nimaimani91@gmail.com

*Abstract*—**In this paper, we investigate the energy efficient power allocation for the downlink in the massive multiple-input multiple-output (MIMO) systems under zero-forcing (ZF) receiver. The radio frequencies (RF) that have a significant effect on system performance has not been considered in most of previous researches. Increasing of random changes in RF creates a mismatch channel. We must consider the effects of the RF circuit for the performance evaluation of the massive MIMO systems. We also consider the rate of system in the presence of estimation error, similar to real world, with the quality of service (QoS) constraint and the transfer capacity of users. In the scenario of this paper, users are divided into two groups. The first group are users who have stronger channel conditions or in other words are located in the center of the cell, and the second group belongs to users who have weak channel conditions and are located at the edge of the cell. By using Karush-Kuhn-Tucker (KKT) conditions, we obtain the optimal power of users. The results of implementation and simulations are given to confirm the efficiency of the proposed algorithm.**

## I. Introduction

Multiple-input multiple-output (MIMO) systems are one of the most effective technologies of the fifth generation (5G) cellular systems. These systems have an acceptable performance in the spectral efficiency (SE) and the energy efficiency (EE). The MIMO systems, in which a base station (BS) is equipped with hundreds of antennas and communicates with tens or hundreds of users simultaneously in a same frequency-time block, are considered. As the number of BS antennas in massive MIMO systems goes to infinite, the noise and dimming effects disappear in the small scale range. Massive MIMO systems provide reliable communication, high EE, and low-complexity schemes for signal processing. Several technologies are involved in the massive MIMO in 5G networks including non-orthogonal multiple access (NOMA), heterogeneous networks, millimeter waves, and device-to-device (D2D) communications [1].

In [2], a closed form solution is proposed using Karush-Kuhn-Tucker (KKT) conditions to maximize the total data rate in the presence of error estimation. In the simulation results of [2], the NOMA and orthogonal multiple access (OMA) scenarios are compared with each other, which shows that NOMA outperforms OMA. It has also been shown that the NOMA interest increases as the number of users within the cluster increases. Most research has been accomplished only to discuss the effects of wireless channels, and it is less common to study the effect of radio frequency (RF) circuits on the system performance. Another closed form expression is presented for detailed analysis of the total data rate in the downlink of the MIMO systems using the maximum ratio transmission (MRT) and zero forcing (ZF) pre-encoding [3]. Furthermore, for the simplest case of user equipment incompatibility with the ZF encoding, the water-filling solution is applied. For other cases of non-conformity, before using some convex optimization solution tools, through convex inequality, variable substitution, and Taylor expansion techniques, a convex relaxation is performed on an asynchronous problem. Finally, an iterative algorithm is proposed to approach the optimal solution.

A multiple-user duplex MIMO system is considered in [4]. In this system, the BS is equipped with a large number of antennas that serve the users of the single antenna that is located around it. In the downlink, it is assumed that users can obtain useful information about the status of the channel with the help of pilot sequences transmitted by the BS to decode the data signals. In this system, each user employs the Minimum Mean Square Error (MMSE) method to obtain estimations of the effective channel gain.

In [5], an energy efficient power allocation for massive MIMO systems by NOMA method is considered. In the proposed scheme, the model system is divided into two areas, the first area is for users whose distance to the BS is less than half of the cell radius, and users whose distance to the BS is more than half of the cell radius are in the second area. A coefficient based on user distance is also introduced, which determines the total amount of the power allocated to each area. Using the standard interference function (SIF), a new iterative algorithm is proposed to solve the optimization problem.

An energy efficient power allocation for the massive MIMO systems with minimum power constraint per user is proposed in [6]. This provides a minimum quality of service (QoS) for each user. In order to maximize the EE, in this article, a minimum power condition has been considered, and if this condition is met, EE will be maximized, otherwise a user is added to the cluster. In this paper, an iterative algorithm is presented, which consistent with the simulation results, it is concluded that its performance is better than the proposed algorithms.

A power limited non-convex optimization problem to maximize the EE of a massive MIMO system is investigated [7]. In this issue, the optimization is required to consider the QoS. Using the fractional programming properties and the lower bound of the sum-rate of users, it is possible to turn the problem into a convex optimization problem. Then, the Lagrange dual function is used to convert a constraint to an unconstraint problem. By using the SIF, an iterative algorithm to solve the optimization problem, i.e., the optimal power allocation scheme, is presented. The simulation results clearly show that the proposed algorithm converges rapidly. Of course, it shows the effects of parameters such as the number of users and the number of antennas on the system performance.

The total data rate resulting from the power allocation in massive MIMO systems with respect to the time division duplex transmission is derived [8]. An energy efficient power allocation scheme for the MIMO-NOMA system is proposed in [9] to reach a minimum acceptable rate for each user by considering multiple users in a cluster. The allocation of efficient energy resources problem for theses systems has not been completely considered and previous work is on the selection of sub-channels and power allocation to maximize the sum rate. The maximization problem of the EE value for the NOMA networks in the downlink are discussed in [10], which uses the optimization techniques and allocating of the sub-channel power. By employing the Newton methods and common user communication based on the Lagrange analysis, an algorithm has been proposed with low-complexity. Based on the results of [11], the proposed algorithm provides maximum EE value for the case that the BS is equipped with a large number of transmitter antennas.

The total data rate is obtained using ZF receivers and pre-encodings in the downlink and uplinks. This paper presents a network-wide scheduling scheme to avoid the joint optimization that is highly complex. The simulation results show that the proposed scheme has a significant advantage in comparison with the equal power allocation.

The main novelties and properties of our proposed scheme in this paper are summarized as follows:

- The effects of RF and channel estimation error are not considered in most research. We include these two factors in the power allocation calculations at the same time and show their effects on system performance.
- Users are divided into two groups. The first group is the users who have strong channel conditions and located in the center of the cell. The second group belongs to the users whose channel conditions are weak and located at the edge of the cell.
- An efficient power allocation algorithm is introduced that calculates the power of users based on the group in which they are located with improved performance.

The structure of this paper is as follows. In Section II, the proposed system model is introduced and the interface is formulated. In Section III, the optimization problem is formulated and then we solve the optimization problem using the proposed solution. In Section IV, the simulation results are presented to evaluate the performance, and finally in Section V, the conclusions of this paper appears.

## II. SYSTEM MODEL

The system model that is considered in this paper is shown in Fig. 1. In this model, users are divided into two categories. The first group are users who are close to BS and the second group are users who are at the edge of the cell. The BS in this system, the model is equipped with an $M$ antenna and users who are randomly distributed around it, with an antenna.

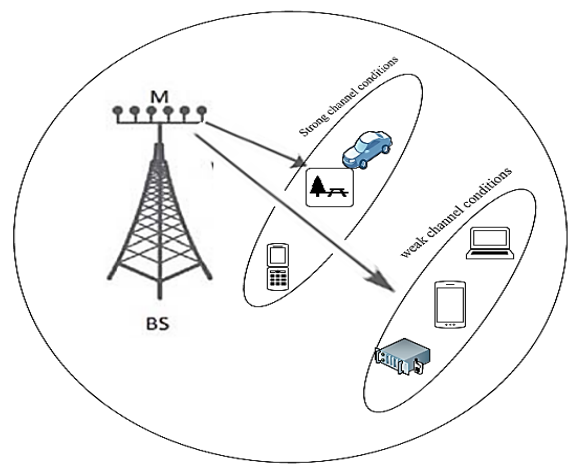

**Fig.1.** System model of the massive MIMO systems.

Assume **G** is the matrix of the flat fading channel between the BS and the $K$ users. **G** can be written as [7]:

$$\mathbf{G} = \mathbf{H}\mathbf{Q}^{1/2}, \quad (1)$$

where, $\mathbf{H} = [\mathbf{h}_1^T, \mathbf{h}_2^T, ..., \mathbf{h}_K^T] \in C^{M \times K}$ is the small scale fading mismatch channel matrix and $\mathbf{Q} = diag\{\beta_1, \beta_2, ..., \beta_K\}$ denotes the large scale fading matrix, where, $\beta_K = \upsilon\varsigma/d_k^\varepsilon$ represents the path loss and shadow fading. $\upsilon$ is a constant related to the carrier frequency and antenna gain, $d_k$ is the distance between the BS and the $k$-th user, $\varepsilon \in [2,6]$ represents the path loss exponent, and $\varsigma$ is the shadow fading with lognormal distribution, i.e., $10\log_{10}(\theta) \sim N(0, \sigma_\varsigma^2)$.

If $\hat{h}_k$ is assumed to be an estimation of $h_k$, the relation $h_k$ can be expressed as $h_k = \hat{h}_k + \epsilon$, where the distribution is $\epsilon \sim \mathcal{CN}(0, \sigma_\epsilon^2)$ [2]. The distribution of $\hat{h}_k$ is a Gaussian distribution with zero mean and variance $\sigma_{\hat{h}_k}^2 = d_k^{-\alpha} - \sigma_\epsilon^2$.

Accordingly, at user $k$, the observed signal is formulated as:

$$y_k = \sqrt{\beta_k p_k \tau_d} \hat{h}_k s_k \frac{u_{r,k}}{u_{t,k}}$$
$$+ \sqrt{\beta_k} \hat{h}_k \sum_{\substack{k'=1 \\ k' \neq k}}^{K} \sqrt{p_{k'} \tau_d} s_{k'} \frac{u_{r,k'}}{u_{t,k'}} \quad (2)$$
$$+ \epsilon \left( \sum_{k=1}^{K} \sqrt{p_k \tau_d} \frac{u_{r,k}}{u_{t,k}} \right) + n_k.$$

$p_k$ denotes the transmission power allocated to the $k$-th user, $h_k$ is the $k$-th column of **G**, $s_k$ represents transmission data symbol of the $k$-th user ($E\{|s_k|^2\}=1$), $\tau_d$ is the length of pilot symbols. $n_k$ shows the additive white Gaussian noise at the $k$-th user with distribution $N(0, \sigma^2)$.

$u_{r,k}$ and $u_{t,k}$ are the complex RF circuit gains, which can be written as $u_{r,k} = |u_{r,k}|e^{i\varphi_{r,k}}$ and $u_{r,t} = |u_{r,t}|e^{i\varphi_{r,t}}$ with $\varphi_{r,k} \sim U[-\theta_{r,k}, \theta_{r,k}]$, where $\theta$ is the maximal mismatch phase of RF gain in the reception of users. The amplitude of the RF gain is assumed to be of log-normal distribution, i.e., $\ln|u_{r,k}| \sim \mathcal{N}(0, \delta_{u,r}^2)$ and $\ln|u_{t,t}| \sim \mathcal{N}(0, \delta_{u,t}^2)$ [3], where, $\delta_{u,k}^2$ is the mismatch variance of the RF gain in the reception of users and $\delta_{u,t}^2$ is the mismatch variance of the RF gain in the transmission of users. The received signal-to-interference-plus-noise ratio (SINR) of the $k$-th user is given by,

$$\gamma_k = \frac{\beta_k p_k \tau_d |u_{r,k}|^2 |u_{t,k}|^{-2} |\hat{h}_k|^2 |s_k|^2}{\tau_d \beta_k \pi + P\tau_d \sigma_\epsilon^2 + \sigma^2}, \quad (3)$$

where,

$$\pi = \sum_{\substack{k'=1 \\ k' \neq k}}^{K} \sqrt{p_{k'}} |u_{r,k}|^2 |u_{t,k'}|^{-2} |\hat{h}_{k'}|^2 |s_{k'}|^2. \quad (4)$$

By considering the length of the pilot symbol ($\tau_d$), and length of coherence interval ($T$) of the downlink transmission, the data rate value is as follows:

$$r_k = \frac{T - \tau_d}{T} \log_2(1 + \gamma_k). \quad (5)$$

The ratio of total user data rates to total user transfer power and fixed circuit power can be introduced as a general definition for the EE. According to this definition we have,

$$EE = \frac{\sum_{k=1}^{K} r_k}{\sum_{k=1}^{K} p_k + \sum_{m=1}^{M} P_{c,m}}, \quad (6)$$

where, $P_{c,m}$ shows the constant power consumed by circuits.

III. THE PROPOSED POWER ALLOCATION ALGORITHM

In this section, we first formulate the proposed energy efficient power allocation scheme and then the method of solving the problem is explained.

*A. Problem Formulation*

In this section, we formulate an energy-efficient power allocation problem for a MIMO system. This optimization problem has three constraints. The purpose of this formulation is to maximize the EE value,

$$\max_{\{p_1, p_2, ..., p_K\}} \eta_{EE} \quad (7.a)$$

$$C1: \sum_{\substack{k=1 \\ k \in K_c}}^{K} p_k \leq \mathcal{k}_c \frac{1}{T - \tau_d} P \quad (7.b)$$

$$C2: \sum_{\substack{k=1 \\ k \in K_e}}^{K} p_k \leq \mathcal{k}_e \frac{1}{T - \tau_d} P \quad (7.c)$$

$$C3: p_k \geq \left(2^{R_k^{min}} - 1\right) \left(\sum_{j=1}^{k-1} p_j + \frac{P\tau_d \sigma_\epsilon^2 + \sigma^2}{\|\hat{h}_k\|^2}\right). \quad (7.d)$$

$P$ denotes the flexible transmission power, $R_k^{min}$ represents the minimum data rate per user. Constraints $C1$ and $C2$ also represent the maximum transfer power constraint (all users' power must be less than a maximum power value) for users with strong channel conditions (set of $K_c$) and users with weak channel conditions (set of $K_e$), respectively. Constraint $C3$ also indicates that each user's transmission power must be greater than a minimum power value. $\mathcal{k}$ indicates how much of each user set (with strong channel conditions and weak channel conditions) utilizes the total transmission power. Its relationship can also be shown as follows:

$$\mathcal{k}_c = \frac{\sum_{j=1}^{K_c} \beta_j}{\sum_{k=1}^{K} \beta_k}, \quad \mathcal{k}_e = (1 - \mathcal{k}_c) = \frac{\sum_{j=1}^{K_e} \beta_j}{\sum_{k=1}^{K} \beta_k}. \quad (8)$$

In this section, the optimization problem was formulated and its constraints were clearly analyzed. In the following, we will solve this optimization problem.

### B. The Proposed Solution

#### 1) Convex transformation

As can be clearly seen from (6), the EE relationship has a fractional form. Therefore, it can be included in the non-convex fraction in the programming classification. By taking advantage of the fractional programming features and the minimum user data rate, the non-convex problem is turned into a convex problem. By following a similar approach as in [12], the maximum EE can be achieved if and only if:

$$\max_{\{p_1,p_2,\ldots,p_K\}} \left\{ \sum_{k=1}^{K} r_k - q^* \left( \sum_{k=1}^{K} p_k + \sum_{m=1}^{M} P_{c.m} \right) \right\} = 0. \quad (9)$$

Using the perfect channel state information features, the Rayleigh fading and Jensen's inequality ($E[\log_2(1 + e^x)] \geq \log_2(1 + e^{E[x]})$), the lower bound of the $k$-th user's data rate is formulated as follows [3]:

$$\tilde{r}_k = \log_2 \left( \frac{\left(|\hat{h}_k|^2 - K\right) \beta_k p_k \tau_d}{e^{2\delta_{u,t}^2 \tau_d \sum_{\substack{k'=1 \\ k' \neq k}}^{K} \beta_{k'}^{-1} p_{k'} + P\tau_d \sigma_\epsilon^2 + \sigma^2}} \right). \quad (10)$$

Let $q$ denote the value of the EE. By using (9) and (10), the optimization problem is reformulated as:

$$\max_{\{p_1,p_2,\ldots,p_K\}} \left\{ \sum_{k=1}^{K} \tilde{r}_k - q^* \left( \sum_{k=1}^{K} p_k + \sum_{m=1}^{M} P_{c.m} \right) \right\} \quad (11.a)$$

$$C1: \sum_{\substack{k=1 \\ k \in K_c}}^{K} p_k \leq \hbar_c \frac{1}{T - \tau_d} P \quad (11.b)$$

$$C2: \sum_{\substack{k=1 \\ k \in K_e}}^{K} p_k \leq \hbar_e \frac{1}{T - \tau_d} P \quad (11.c)$$

$$C3: p_k \geq \left(2^{R_k^{min}} - 1\right) \left( \sum_{j=1}^{k-1} p_j + \frac{P\tau_d \sigma_\epsilon^2 + \sigma^2}{\|\hat{h}_k\|^2} \right). \quad (11.d)$$

The obtained relation is a simplified which is a convex optimization problem. Now, we use the Lagrangian dual function to turn this problem into a unconstraint convex problem.

#### 2) Elimination of constraints

Considering $\Phi$ as the dual Lagrangian function, problem (11) is simplified as follows [13]:

$$\Phi(\mathcal{P}, \omega_c, \omega_e, \lambda_k) = \quad (12)$$
$$-\omega_c \left[ \hbar_c \frac{1}{T-\tau_d} P \sum_{\substack{k=1 \\ k \in K_c}}^{K} p_k \right] - \omega_e \left[ \hbar_e \frac{1}{T-\tau_d} P - \sum_{\substack{k=1 \\ k \in K_e}}^{K} p_k \right] - \lambda_k \left( p_k - \left(2^{R_k^{min}} - 1\right) \left( \sum_{j=1}^{k-1} p_j + \frac{P\tau_d \sigma_\epsilon^2 + \sigma^2}{\|\hat{h}_k\|^2} \right) \right),$$

where, the parameter $\mathcal{P}$ represents the acceptable power for each user. The parameters $\omega_c \geq 0$ and $\omega_e \geq 0$ represent the Lagrangian multiplier corresponding to the transmission power constraint for users with strong and weak channel conditions, respectively. Furthermore, $\lambda_k \geq 0$ indicates the Lagrangian multiplier corresponding to the minimum rate required constraint by users.

#### 3) Iterative Algorithm

In order to obtain the optimal allocated power for users, it is necessary to derive the Lagrangian dual function and then set its derivation to zero. This should be done for both groups of users.

If $\beta_k \geq \ell$, i.e., users whose channel conditions are greater than a certain value, the following relation is used. Let $\ell$ denotes the predetermined positive threshold for the path loss and shadow fading ($\beta$).

$$\frac{\partial \phi}{\partial p_k} =$$
$$\sum_{j=1, j \neq k} \frac{e^{2\delta_{u,t}^2 \tau_d \sum_{\substack{k'=1 \\ k' \neq k}}^{K} \beta_{k'}^{-1}}}{\left( e^{2\delta_{u,t}^2 \tau_d \sum_{\substack{k'=1 \\ k' \neq k}}^{K} \beta_{k'}^{-1} p_{k'} + P\tau_d \sigma_\epsilon^2 + \sigma^2} \right) \ln 2}$$
$$-\frac{1}{p_k \ln 2} + q + \omega_c + \left(\lambda_k - \sum_{j=1}^{k-1}(\omega_j - 1)\lambda_j\right) = 0. \quad (13)$$

If $\beta_k < \ell$, i.e., users whose channel conditions are less than a certain value, the following relation is used.

$$\frac{\partial \phi}{\partial p_k} =$$
$$\sum_{j=1, j \neq k} \frac{e^{2\delta_{u,t}^2 \tau_d \sum_{\substack{k'=1 \\ k' \neq k}}^{K} \beta_{k'}^{-1}}}{\left( e^{2\delta_{u,t}^2 \tau_d \sum_{\substack{k'=1 \\ k' \neq k}}^{K} \beta_{k'}^{-1} p_{k'} + P\tau_d \sigma_\epsilon^2 + \sigma^2} \right) \ln 2}$$
$$-\frac{1}{p_k \ln 2} + q + \omega_e + \left(\lambda_k - \sum_{j=1}^{k-1}(\omega_j - 1)\lambda_j\right) = 0. \quad (14)$$

According to (13), the optimal power for users with strong channel conditions is formulated as follows:

$$p_k = \frac{1}{\left( \sum_{j=1, j \neq k} \frac{\Omega}{(\Lambda \ln 2) \ln 2} + \omega_c + q + \chi \right) \ln 2}, \quad (15)$$

where,

$$\Omega = e^{2\delta_{u,t}^2}\tau_d \sum_{\substack{k'=1 \\ k'\neq k}}^{K} \beta_{k'}^{-1}, \tag{16}$$

$$\Lambda = e^{2\delta_{u,t}^2}\tau_d \sum_{\substack{k'=1 \\ k'\neq k}}^{K} \beta_{k'}^{-1} p_{k'} + P\tau_d \sigma_\epsilon^2 + \sigma^2, \tag{17}$$

$$\chi = \left(\lambda_k - \sum_{j=1}^{k-1}(\omega_j - 1)\lambda_j\right). \tag{18}$$

Similarly, the optimal power for users with weak channel conditions is formulated as follows:

$$p_k = \frac{1}{\left(\sum_{j=1,\ j\neq k} \frac{\Omega}{(\Lambda\ ln2)ln2} + \omega_e + q + \chi\right)\ln 2}. \tag{19}$$

In the following, we provide our proposed algorithm in a step by step procedure. Let $n$ denote the iteration number. $\varpi_1$, $\varpi_2$ and $\Theta_k$ denote the positive step sizes in the algorithm.

---

**Algorithm 1** Energy efficient power allocation algorithm for the massive MIMO system.

**Step 1**: Initialize
$R_k^{min}$, $u = 0$, $v = \frac{\zeta}{\sigma^2\ ln2} \min_k \{\beta_k \|\hat{\mathbf{h}}_k\|^2\}$, $\varepsilon > 0$ $p_k > 0$ ($k = 1,2,\ldots,K$) transmission power and Lagrangian multipliers P, $\lambda_k^{(0)}, \omega_c^{(0)}, \omega_e^{(0)}$

**Step 2**: Calculate
Let $\eta_{EE} = \frac{u+v}{2}$,
If ($\beta_k \geq \ell$):
  solve $p_k$ according to formula (15),
else if ($\beta_k < \ell$):
  solve $p_k$ according to formula (19),
end if
Then a realistic EE $q$ can be got by formula (5),
  If_2 ($\eta_{EE} \geq EE$), then $u = \eta_{EE}$
  Else $v = \eta_{EE}$.
  End if_2

**Step 3: Update**

$$\omega_c^{(n+1)} = max\left(0, \omega_c^{(n)} - \varpi_1\left(k_c \frac{1}{T-\tau_d}P - \sum_{\substack{k=1 \\ k\in K_c}}^{K} p_k^{(n)}\right)\right)$$

$$\omega_c^{(n+1)} = max\left(0, \omega_c^{(n)} - \varpi_2\left(k_e \frac{1}{T-\tau_d}P - \sum_{\substack{k=1 \\ k\in K_e}}^{K} p_k^{(n)}\right)\right)$$

$$\lambda^{(n+1)} = max(0, \lambda^{(n)} - \Theta_k(p_k - \left(2^{R_k^{min}} - 1\right)\left(\sum_{j=1, j\neq k}^{K} p_j + \frac{P\tau_d\sigma_\epsilon^2 + \sigma^2}{\|\hat{\mathbf{h}}_k\|^2}\right))$$

**Step 4: Optimal solution**
if_3 ($v - u < \varepsilon$) then $q \approx \frac{u+v}{2}$
 Else go to Step 2.
 End if_3
 n=n+1
**end**

---

TABLE I. The simulation parameters.

| Parameters | Value |
|---|---|
| Noise spectral density $N_0$ | -170 dBm/Hz |
| Variance of log-normal shadow fading $\sigma^2$ | 10 dB |
| Efficiency of PA ($\zeta$) | 0.5 |
| Factor $v$ | 1 |

## IV. SIMULATION RESULTS

In this section, we analyze the simulation results. A total transmission power of 1W is considered. In the system model the radius of cell is assumed to be 500 meters. The important parameters of this system that are used in the simulation process are given in Table 1.

Figure 2 shows the EE performance of the system with increasing the RF mismatch variance. The channel estimation error for this figure is equal to 0.03. This figure is drawn for two scenarios ($M = 200$ and $M = 100$). Obviously, with increasing the RF variance, the EE of the system decreases. But there are two very important points in this figure. First, the algorithm proposed in this paper performs better than equal power allocation, and second, when the RF variance is larger ($\delta_{u,r}^2 = \delta_{u,t}^2 > 2$), the large number of antennas does not improve system performance and gives almost the same results. In other words, the proposed algorithms are effective in this type of system model for a low amount of RF mismatch variance.

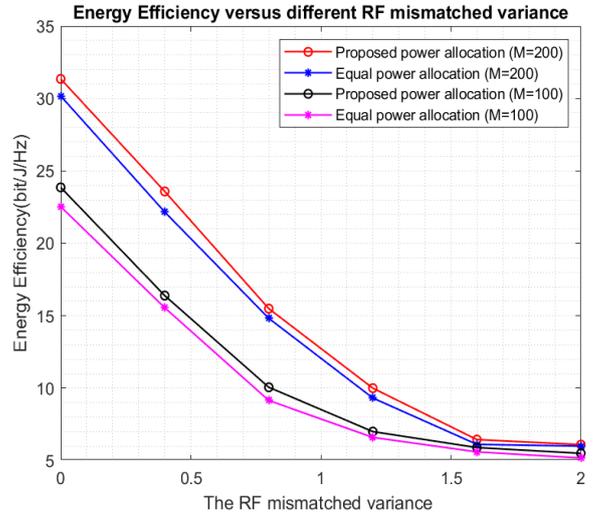

**Fig. 2.** EE versus different RF mismatched variance ($\delta_{u,r}^2 = \delta_{u,t}^2$) ($P = 20$ dBm).

Figure 3 shows the performance of the total system data rate with increasing of the channel estimation error. In this figure, two scenarios of two-user and three-user are presented. The RF mismatch variance is assumed to be 0.3. Two scenarios for $\sigma_\epsilon^2 < 0.05$ have almost the same performance, and for larger values the performance improvement of the proposed algorithm is clearly seen. For instance, for the two-user case, at $\sigma_\epsilon^2$=0.2, the value of the improvement in total data rate is about 0.26 bit/s/Hz, and this gap value is about 0.27 bit/s/Hz for the three-user case.

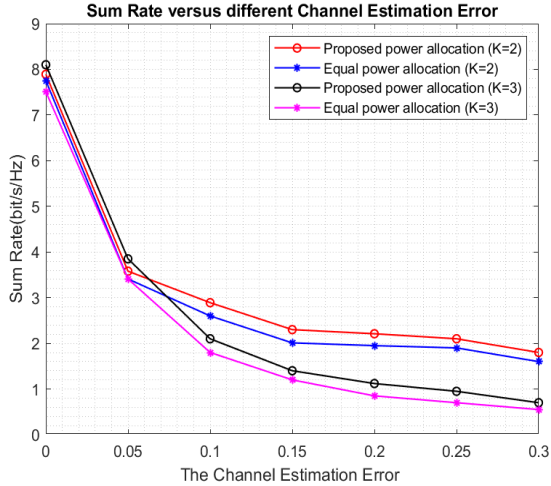

**Fig. 3.** SE versus different channel estimation error variance($\sigma_\epsilon^2$). ($P$ = 20 dBm).

In Table II, we show the EE performance variations for some values of the minimum user rate. Furthermore, the amount of improvement in the proposed algorithm is presented in comparison with the equal power allocation scheme. According to the obtained results, by increasing the minimum user rate, the results of two methods get closer to each other.

TABLE II. The EE versus different values of $R_k^{min}$.

| User rate constraint ($R_k^{min}$) | EE convergence in the proposed algorithm (bit/J/Hz) | EE convergence algorithm in the equal power allocation method (bit/J/Hz) | Amount of improvement (bit/J/Hz) |
|---|---|---|---|
| 1 | 30.478 | 28.947 | 1.531 |
| 1.5 | 28.465 | 27.296 | 1.169 |
| 2 | 24.368 | 23.412 | 0.956 |

## V. CONCLUSION

In this paper, the allocation of energy efficient power for a massive MIMO system is investigated. Effective parameters, namely the radio frequencies and the channel estimation error, which have received less attention in most researches, are considered in this paper and their effect on system parameters is shown. Users are divided into two groups. The first group included users with strong channel conditions and the second group included users with poor channel conditions. An efficient power allocation algorithm is presented and the implementation results confirms that it outperforms the equal power allocation scheme.